# β-delayed proton decay of proton-rich nuclei $^{23}$Al and $^{31}$Cl and explosive H-burning in classical novae

L. Trache[1], A. Banu[1], J.C. Hardy[1], V.E. Iacob[1,*], M. McCleskey[1], E. Simmons[1],

G. Tabacaru[1,*], R.E. Tribble[1], J. Aysto[2], A. Jokinen[2], A. Saastamoinen[2],

M.A. Bentley[3], D. Jenkins[3], T. Davinson[4], P.J. Woods[4], N.L. Achouri[5], B. Roeder[1,5]

[1]*Cyclotron Institute, Texas A&M University, College Station, TX 77843, USA*
E-mail: livius_trache@comp.tamu.edu
[2]*University of Jyvaskyla, Jyvaskyla, Finland*
[3]*University of York, York, UK*
[4]*University of Edinburgh, Edinburgh, UK*
[5]*LPC, ENSICAEN, University of Caen, CNRS/ IN2P3, Caen, France*

We have developed a technique to measure beta-delayed proton decay of proton-rich nuclei produced and separated with the MARS recoil spectrometer of Texas A&M University. The short-lived radioactive species are produced in-flight, separated, then slowed down (from about 40 MeV/u) and implanted in the middle of very thin Si detectors. The beam is pulsed and βp decay of the pure sources collected in beam is measured between beam pulses. Implantation avoids the problems with detector windows and allows us to measure protons with energies as low as 200 keV from nuclei with lifetimes of 100 ms or less. Using this technique, we have studied the isotopes $^{23}$Al and $^{31}$Cl, both important for understanding explosive H-burning in novae. They were produced in the reactions $^{24}$Mg(p,2n)$^{23}$Al and $^{32}$S(p,2n)$^{31}$Cl, respectively, in inverse kinematics, from stable beams at 48 and 40 MeV/u, respectively. We give details about the technique, its performances and the results for $^{23}$Al and $^{31}$Cl βp-decay. The technique has shown a remarkable selectivity to β-delayed charged-particle emission and would work even at radioactive beam rates of a few pps. The states populated are resonances for the radiative proton capture reactions $^{22}$Na(p,γ)$^{23}$Mg and $^{30}$P(p,γ)$^{31}$S, respectively.

---

[*] On leave from IFIN-HH Bucharest, Romania

## 1. Introduction

Classical novae are relatively common events in our Galaxy, with models predicting about 30 yr$^{-1}$, but with only a few per year actually being detected. This large frequency of appearance makes them easier to study by observations in various parts of the electromagnetic spectrum, and they are targets of many models. The current understanding is that they occur in interacting binary systems where H-rich material accretes on a white dwarf from its low-mass main-sequence companion. At some point in the accretion the H-rich matter compresses and leads to a thermonuclear runaway (TNR), during which nucleosynthesis occurs [1]. Once the dynamics of nova outbursts and the nucleosynthesis fueling it are understood, their contribution to the chemical evolution of the Galaxy can be better assessed through the comparison of predictions with observations. Novae are anticipated to become the first type of explosive phenomena where all nuclear data for nucleosynthesis can be based on experimental data [2]. However, we are still far from reaching that goal. Among the key reactions for which the reaction rates are known with large uncertainties are the radiative proton captures $^{22}$Na(p,γ)$^{23}$Mg and $^{30}$P(p,γ)$^{31}$S. These reaction rates are dominated by capture through low-energy proton resonances, which correspond to excited states in $^{23}$Mg and $^{31}$S nuclei. Considerable efforts were and are being made to find these resonances and to determine their parameters (position and resonance strength) by direct or indirect methods. Some of these experiments are presented at this symposium [3]. We adopt an approach involving the study of excited states in $^{23}$Mg and $^{31}$S via β-decay of their parent nuclei, $^{23}$Al and $^{31}$Cl, respectively. States populated above the proton binding energy in each nucleus can decay by proton emission (β-delayed proton decay, or βp) and are the resonances we are seeking.

## 2. The experimental technique. βp-decay of $^{23}$Al

We have developed a technique to measure β-delayed proton decay of proton-rich nuclei produced and separated with the MARS recoil spectrometer at the K500 superconducting cyclotron of Texas A&M University. The short-lived radioactive species are produced in-flight, separated, then slowed down (from about 40 MeV/u) and implanted in the middle of a very thin Si detector. Implantation avoids the problems with detector windows or dead layers and allows us to measure protons with energies as low as 200 keV from nuclei with lifetimes of 100 ms or less. The technique provides a valuable tool for the study of states that are resonances important in the radiative proton capture on nuclei close to the proton drip line. It also gives information on Isobaric Analog States (IAS) and possible isospin mixing.

Using this technique, we have studied the isotopes $^{23}$Al and $^{31}$Cl. The sources were produced in-flight through the reactions $^{24}$Mg(p,2n)$^{23}$Al and $^{32}$S(p,2n)$^{31}$Cl, in inverse kinematics, from stable beams at 48 and 40 MeV/u, respectively. We have studied $^{23}$Al β-decay before [4,5] using β–γ coincidence techniques. Secondary beam rates of about 4000 pps and >90% purity were obtained. The states populated in $^{23}$Mg above the proton threshold at S$_p$=7580 keV can decay by proton emission. They are resonances in the proton capture reaction $^{22}$Na(p,γ)$^{23}$Mg, crucially important for the depletion of $^{22}$Na in ONe novae. A setup consisting of a thin Si double-sided strip detector (p-detector, 65 μm, 16x16 strips) and a thick Si detector (β-detector, 1 mm) was used in the present experiment. A HPGe detector outside the chamber has detected the γ-rays. We have pulsed the beam from the cyclotron, implanting the source nuclei in the thin Si detector (for about 2 lifetimes), and then switched the beam off (same duration) and measured β–p and β–γ coincidences. In order to reduce to a minimum and control the implantation depth we have restricted the momentum spread of the incoming $^{23}$Al nuclei to about ±0.25% and the beam rate to about 500 pps. This has been done by closing down the momentum defining slits in MARS. The implantation depth was controlled using a rotating energy-degrader foil in front of the Si telescope. Implantation distributions of the order of 17 μm deep were obtained. After β-decay either gamma or proton decay follows. All protons emitted with energies below 1.5 MeV stop in the thin Si strip detector and give sharp peaks. The positrons (emitted before) leave a small signal of continuum spectrum in the same thin detector that adds to the proton signal to produce a skewing of the proton peaks on the high energy side (Fig.1) and degrade the resolution. For those (majority) cases where gamma rays are emitted instead, the positrons give a large background at low energies in the p-detector. To reduce this background and the degradation of the resolution it is essential to make the proton detector as thin as possible and the volume of the detector as small as possible (narrow strips). Almost half of the emitted positrons end up in the thick detector behind and trigger the acquisition when a gamma ray or a proton signal arrives in coincidence. Data on the β-delayed proton decay of $^{23}$Al existed before, but were obtained with less intense sources and at times were contradictory [6, 7]. Detailed results of the present study will be published elsewhere [8]. One remarkable result of the current measurement was obtained when we calibrated our proton detector. Along with $^{23}$Al, $^{21}$Mg, a known β-delayed proton emitter, was produced at a much lower rate. By changing the tilting angle of the energy degrading foil, we could completely stop $^{23}$Al in it, while implanting $^{21}$Mg in the middle of the p-detector. With a production rate of about 1 pps, we could obtain a reasonable calibration spectrum in about 8 hrs of experiment. This shows the sensitivity and selectivity of our method.

### 3.    β and βp-decay of $^{31}$Cl

The βp-decay of $^{31}$Cl was studied in a similar way. A pure $^{31}$Cl source was obtained for the first time at a maximum rate of about 3000 pps and >90% purity. Its β-decay populates highly excited states in $^{31}$S, which may then p-decay or γ-decay. The observed proton lines give the position of resonant states which contribute most to the $^{30}$P(p,γ)$^{31}$S reaction, a critical point in explosive H-burning in novae. At the time of the study, the uncertainty in the reaction rate was estimated to be about a factor 100 [9]. Figure 1 shows a proton spectrum from the β-delayed p-decay of $^{31}$Cl. At the same time a clean gamma-ray spectrum from the βγ-decay of $^{31}$Cl was obtained for the first time with good statistics. From it, we could establish the beta-decay scheme shown in Figure 2. Energy calibration of the gamma-ray spectrum up to about 7 MeV has been made using lines from the decay of $^{32}$Cl produced and separated in the same run. Previously little information was known about the βγ-decay of $^{31}$Cl [10]. We could establish the position and decay of the T=3/2 isobaric analog state of the $^{31}$Cl g.s. in $^{31}$S through four decay paths to the g.s. We obtained $E_{exc}$=6279.5±0.3(stat)±1.5(syst) keV. This precise determination of the IAS energy allowed us to determine more precisely the mass excess ME($^{31}$Cl)= -7,064(8) keV by using the Isobaric Multiplet Mass Equation (IMME) and, therefore, the $Q_{EC}$=11980(8) keV value shown in Fig. 2.

The calibration of the proton spectra was made using known lines from the decay of the proton emitter $^{29}$S, separated in the same run. A situation similar to that of $^{21}$Mg mentioned for the $^{23}$Al measurement was found (with a rate of a few pps we could obtain good calibration spectra in about 8 hrs of beamtime). All data should be regarded as preliminary at this point.

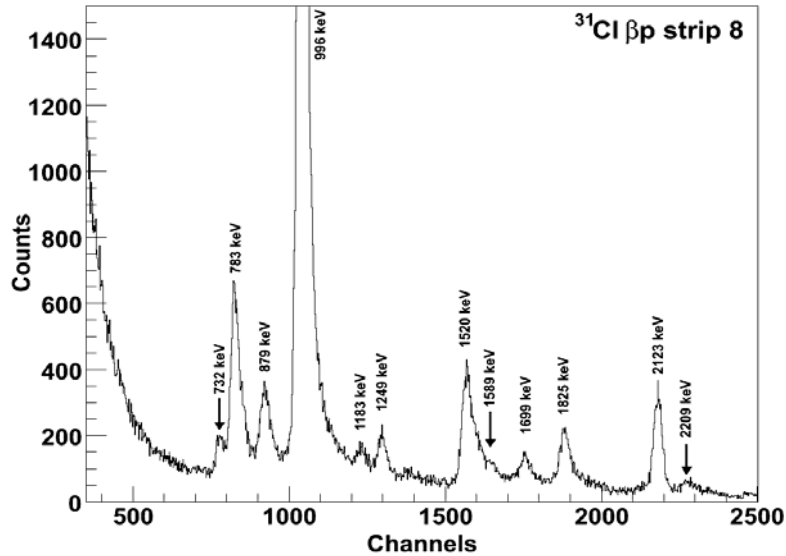

Figure 1: Proton spectrum from the decay of $^{31}$Cl. The uncertainty of proton energies is about 10 keV (mostly systematic).

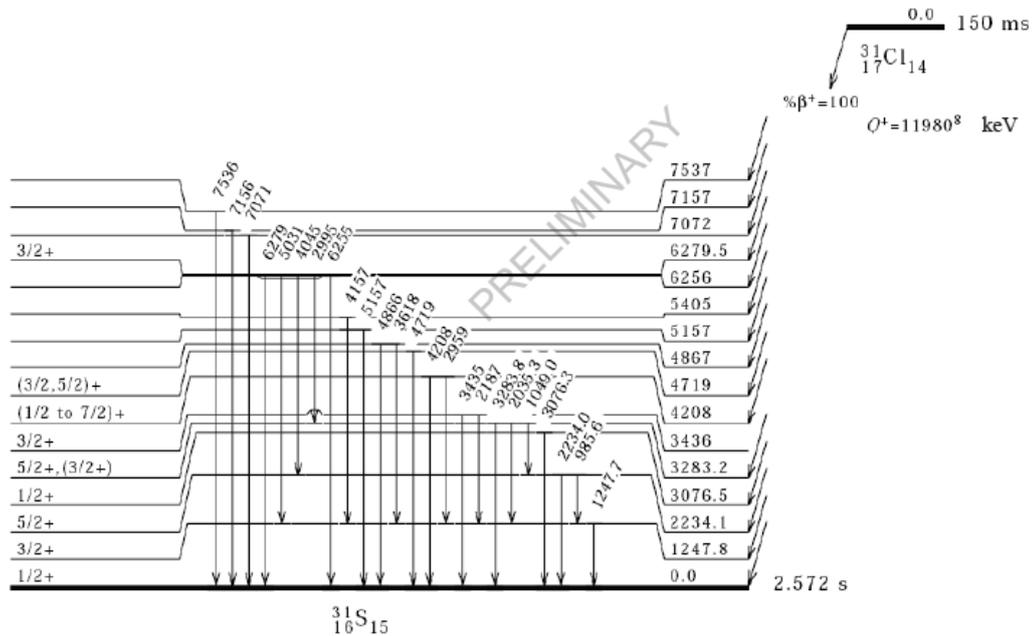

Figure 2: Beta-decay of $^{31}$Cl. Only the levels seen in βγ-decay are included here.

## 4. Conclusions

We have used a technique to measure β-delayed proton decay of proton-rich nuclei produced and separated in-flight with the MARS recoil spectrometer of Texas A&M University. The short-lived radioactive species are implanted in the middle of a thin (65 μm) Si double sided strip detector to detect protons, backed by a thick Si detector for coincident positrons. The beam was pulsed and βp-decay of pure sources was measured between beam pulses. A HPGe detector placed outside the chamber measured simultaneously βγ-coincidences. The technique has shown a remarkable selectivity to β-delayed charged particle emission and would work even at radioactive beam rates of a few pps. Beta-delayed proton decay of $^{23}$Al and $^{31}$Cl was measured with this technique. New resonant states were identified in the region of the Gamow peaks at temperatures relevant for novae and the energies were better determined for a few previously known resonances.